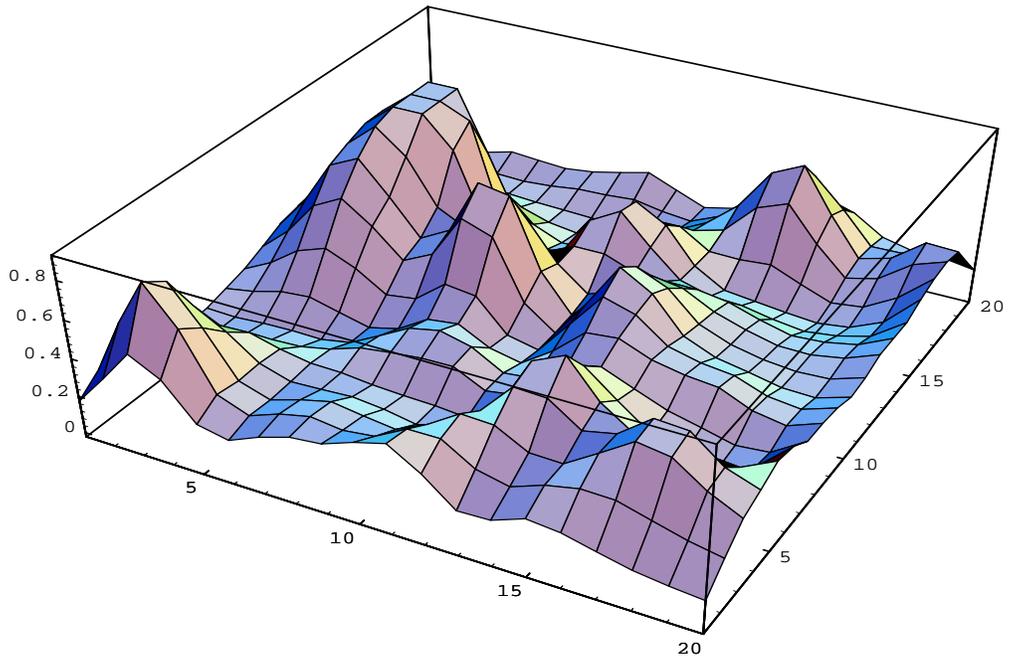

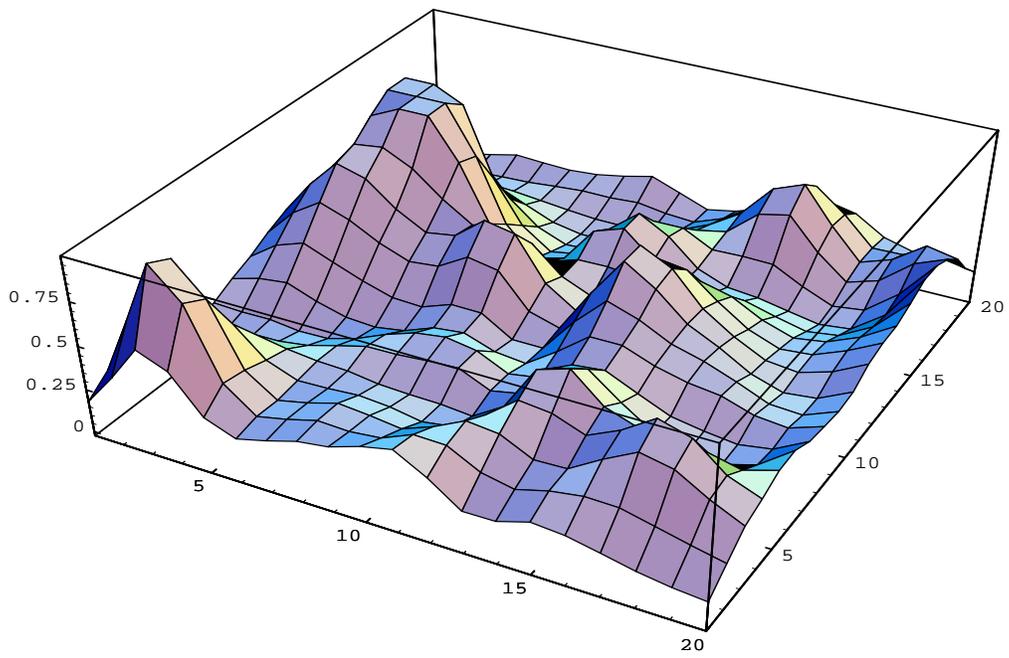

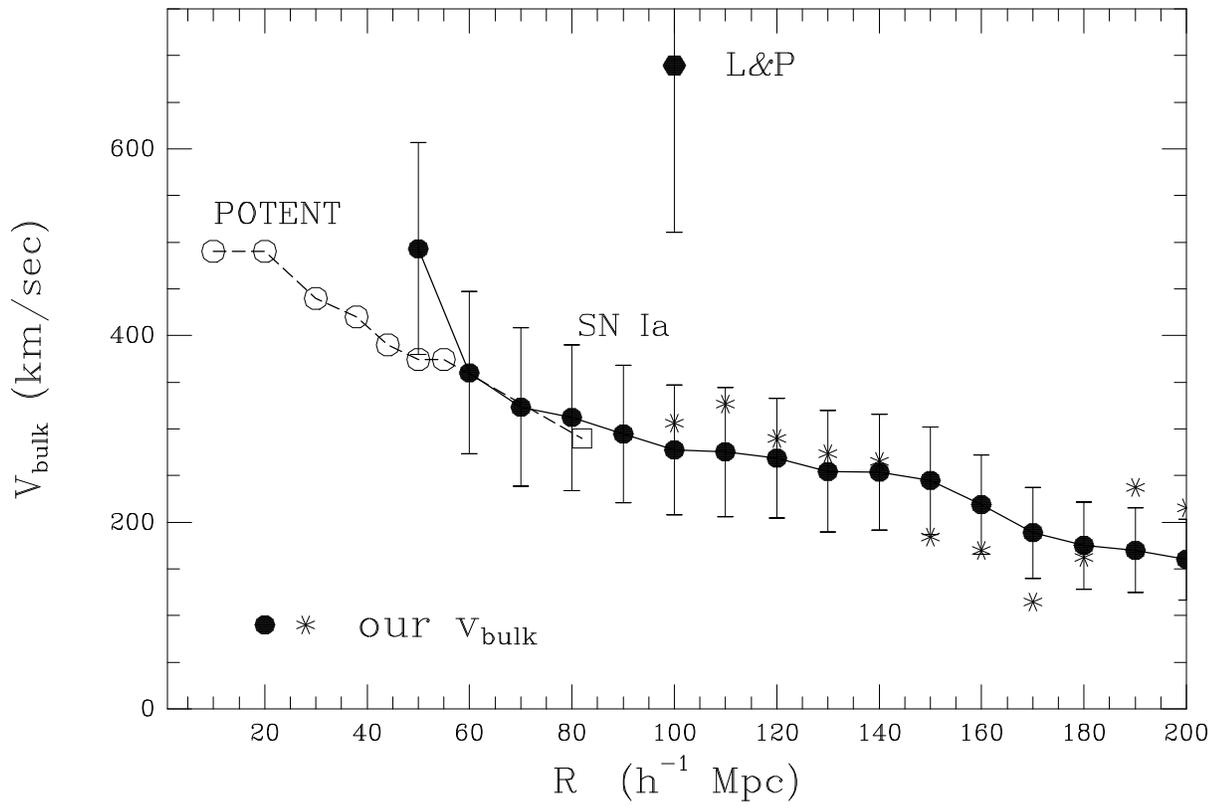

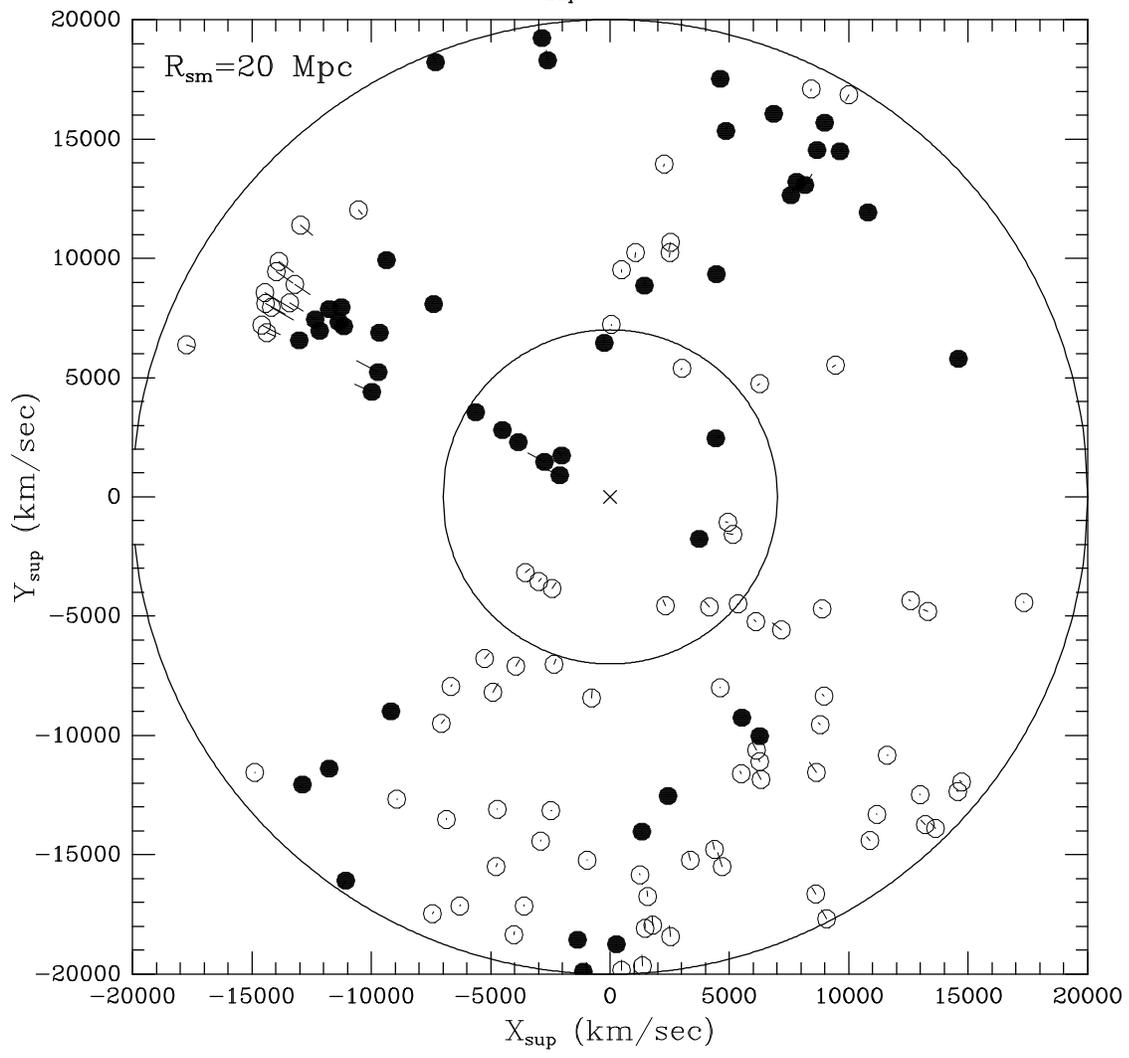

# RECONSTRUCTING POSITIONS AND PECULIAR VELOCITIES OF GALAXY CLUSTERS WITHIN 20000 Km/s

Enzo BRANCHINI[1,2] and Manolis PLIONIS[1,3]
[1] *International School for Advanced Studies, Via Beirut 2-4 34014 Trieste, Italy*
[2] *Queen Mary and Westfield College, Mile End Road, London E1 4NS*
[3] *National Observatory of Athens - Astronomical Institute, Lofos Nimfon, Thesio, 11810, Athens, Greece*

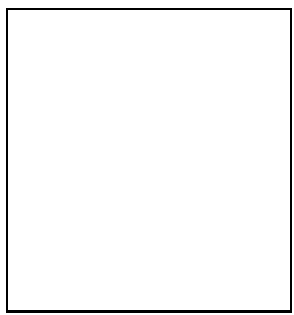

**Abstract**

Starting from the observed redshift space distribution of a volume limited sample of Abell+ACO clusters we use a two step reconstruction procedure to recover their distances and peculiar velocities. The resulting cluster 3-D dipole, as the $z$-space one, converges at $\sim 170\ h^{-1}$ Mpc and points $\sim 10°$ away from the CMB apex. However, redshift space distortions cause the asymptotic dipole of the $z$-space cluster distribution to be overestimated with respect to the 3-D one by $\sim 23\%$, leading to a value of $\beta(\equiv \Omega_0^{0.6}/b) \approx 0.21$, where $b$ is the bias factor of the cluster population with respect to the underline matter distribution.

The resulting cluster velocity field is dominated by a large scale streaming motion along the Perseus Pisces–Great Attractor base-line directed towards the Shapley concentration, in qualitative agreement with the galaxy velocity field on smaller scales. The cluster bulk velocity agrees extremely well with that deduced by the POTENT analysis of galaxy motions at $\sim 50 - 60\ h^{-1}$ Mpc; it decreases thereof while pointing in the general direction of the Cosmic Microwave Background (CMB) dipole, in strong disagreement with the Lauer & Postman (1994) result.

## 1 The Reconstruction Method

We considered the redshift space distribution of the Abell+ACO clusters in the following region:
– Abell: $d \leq 250\ h^{-1}$ Mpc, $|b| > 20°$. N=250 clusters. $m_{10} \leq 17$.
– ACO: $d \leq 250\ h^{-1}$ Mpc, $|b| > 20°$, $m_{10} \leq 17$. N=205 clusters.
– Abell+ACO: $d \leq 250\ h^{-1}$ Mpc, $13° < |b| < 20°$, $m_{10} \leq 17$. N=20 clusters.
98% of clusters have measured redshift. For the remaining objects the redshift is estimated using the $m_{10} - z$ relation, of [4], where $m_{10}$ is the magnitude of the tenth brightest galaxy in the cluster.

We then use a two step procedure to reconstruct the 3-D cluster distribution starting from their observed redshift space positions:

(1) The first step consists in reconstructing the whole sky redshift-space cluster distribution. This is done by generated a population of synthetic objects outside the zone of avoidance (i.e. at $|b| > 20°$) and within 20000 km/s. We impose the synthetic objects to be correlated with the real ones in order to reproduce the observed spatial cluster–cluster correlation function while their large scale distribution accounts for galactic absorption, radial selection function and inhomogeneities between the Abell and ACO cluster catalogues. The object distribution within the zone of avoidance is then recovered by randomly cloning the cluster distribution within redshift–longitude bins in the nearby galactic strips. The reconstruction within the [20000, 25000] km/s region, were the radial selection function falls below unity, is aimed at providing the boundary condition for the next step. In this region the synthetic clusters are distributed according to the cluster selection functions but randomly with respect to the real ones to alleviate the shot noise effects related with the exponential decrease of the real cluster number density. The resulting cluster distribution, however, is a considerable improvement over the simple Poissonian one. Finally, the density field beyond 25000 km/s is considered homogeneous and isotropic.

(2) In the second step we use an iterative reconstruction technique similar to that proposed by [5] to reconstruct the whole sky cluster 3-D distribution and their peculiar velocities. The reconstruction algorithm relies on the linear gravitational instability hypothesis and assumes linear biasing. Tests aimed at estimating the intrinsic reliability of the method showed that the volume average mean error in the reconstructed cluster positions (and velocities) was $< 1.6\%$ [1], for details). Since the whole reconstruction procedure is based on Monte Carlo realizations of the synthetic cluster distribution, all the analyses described below were performed by averaging over several different realizations of the synthetic cluster population.

## 2  The Density Field and the Cluster Dipole

Our reconstruction procedure allows us to recover the true 3-D positions of clusters by removing their redshift space distortions. This can be fully appreciated by comparing the $z$-space and 3-D space reconstructed density fields. In figure 1 we plot a 3-D visualization of the $z$-space density field (left panel) and the 3-D one (right panel) for a slice of 8000 km/sec wide, centered on the supergalactic plane. It is evident that eliminating the redshift distortions reduces significantly the amplitude of the peaks in $z$-space with respect to that in the 3-D space. The effect is particularly evident in the Great Attractor region [(X,Y)=(8,11)], while in the Perseus-Pisces region [(X,Y)=(13,10)] the opposite is true, due probably to a negative velocity gradient in the P-P region; increasing towards the LG.

Figure 1: Surface plot visualization of the density field of a slice 8000 km/sec wide centered onto the supergalactic plane both in z space (left panel) and 3-D space (right panel).

Changing the relative amplitude of the Great Attractor and Perseus-Pisces density peaks has a non negligible effect on the 3-D cluster dipole: redshift space distortions increase the measured dipole amplitude by $\sim 23\%$ leading to an overestimate of the $\beta$ parameter. The direction of the dipole, however, that points $\sim 10°$ away from the CMB apex, does not change appreciably. Comparing the cluster dipole with that of the CMB provides an estimate of $\beta$:

$$\beta = 0.21^{+0.07}_{-0.10} \left[ \frac{622 - V_{in} \cos \delta\theta}{622 - 180 \cos \delta\theta} \right] \qquad (1)$$

where $\delta\theta$ is the misalignment angle between the directions of the Virgo cluster and the CMB dipole ($\delta\theta \sim 45°$) and $V_{in}$ is the Virgocentric infall velocity of the Local Group, the introduction of which is necessary because the contribution to the cluster dipole of the Virgo Abell-like cluster is neglected since, due to its proximity, it is not included in the Abell/ACO sample. Note that the quoted uncertainty is dominated ($\sim 80\%$) by our estimate of the intrinsic cosmic variance (see [1] for details). This result is consistent with an $\Omega_o = 1$ universe for $b = 4.8$.

## 3 The Peculiar Velocity Field and the Bulk Flow

The reconstructed velocities of clusters allow us to explore the dynamics on scales much larger than those probed by luminous galaxies. Because of the low spatial density, however, galaxy clusters provide a sparse sampling of the cosmic velocity field and does not allow us to characterize the velocity field with a resolution comparable to that obtained from galaxy samples. Nevertheless we are able to visualize large scale features of the linear velocity field since the typical errors of the 1-D cluster peculiar velocities are typically $< 150$ km/sec, within the volume sampled.

In Figure 2 we present the cluster peculiar velocity field in a 8000 km/sec wide slice projected onto the supergalactic plane, where most prominent superclusters lie. The velocity field for a Gaussian smoothing of the density field with $R_{sm} = 20\ h^{-1}$ Mpc is shown. Open and filled dots refer to inflowing and outflowing objects, respectively while the length of each vector is equal to to the line of sight component of the peculiar velocity in the CMB frame projected onto the supergalactic plane. The small circle at the center represents the typical region spanned by dynamical analyses based on galaxy peculiar velocities. The most prominent feature is a large coherent motion in the general direction of the CMB dipole towards the Shapley Concentration $(X, Y) = (-13000, +9000)$ km/s. Besides the extent of the bulk flow, in agreement with galaxy studies on smaller scales but extending even beyond those scales, we found again in agreement with galaxy studies, that the reconstructed line of sight component of the cluster peculiar velocities in the Coma region are very small, and that the large-scale coherent motion does not have a constant amplitude; it is small in the Perseus-Pisces region then rises in the Great Attractor zone while dropping on the backside of the Great Attractor. Furthermore, due to our large volume sampled we are able to identify other important features of the large-scale peculiar velocity field not previously studied. For example, it is evident that the bulk velocity rises again near the Shapley concentration where a large back infall is also apparent.

Fig. 2 The cluster peculiar velociy field in a 8000 km/sec wide slice projected onto the supergalactic

plane.

We also computed the bulk velocity using two semi-independent techniques: by computing the center of mass velocity of the region centered on the LG and also by estimating the residual velocity of the whole cluster frame, based on a minimization procedure of the line-of-sight components of the cluster velocities. In figure 3 we plot both estimators as filled dots and starred symbols, respectively. Error bars represent 1 $\sigma$ uncertainties. Although the two estimators are very different they provide a consistent estimate of the bulk velocity, well within their errors. We also plot as open dots the bulk flow obtained by the POTENT reconstruction of the velocity field traced by Mark III galaxies [2].Although the region where both, the galaxy and cluster bulk velocity estimates, overlap is quite small (between $\sim$ 50 and 60 $h^{-1}$ Mpc), they appear to be consistent with each other. Finally, we point out that the bulk flow at $\sim$ 100 $h^{-1}$ Mpc ([3] sample effective limit) has an amplitude of $\sim$ 250 km/sec, much smaller than that found by [3] ($\sim$ 700 km/sec) and thus in comfortable agreement with most cosmological models.

Fig. 3 Cumulative bulk flow from clusters (filled dots and starred symbols), POTENT analysis (open dots) and the [3] result (filled exagon).